\documentclass[preprint,showpacs,preprintnumbers,amsmath,amssymb]{revtex4}

\usepackage{graphicx}
\usepackage{bm}

\newcommand{\E}[1]{Eq.~(\ref{#1})}
\newcommand{\F}[1]{Fig.~\ref{fig:#1}}

\begin{document}

\preprint{nlin.CD/000000}

\title{Bifurcation scenario to Nikolaevskii turbulence in small systems} 

\author{Dan Tanaka}
 \email{dan@ton.scphys.kyoto-u.ac.jp}
\affiliation{%
Department of Physics, Graduate School of Sciences, Kyoto University, Kyoto 606-8502, Japan
}%
\date{\today}%

\begin{abstract}   
We show that 
the chaos in Kuramoto-Sivashinsky equation occurs 
through period-doubling cascade (Feigenbaum scenario),  
in contrast, 
the chaos in Nikolaevskii equation occurs 
through torus-doubling bifurcation (Ruelle-Takens-Newhouse scenario). 
\end{abstract}


\maketitle
We concern the onset of spatiotemporal chaos exhibited by 
Nikolaevskii equation \cite{onsetNik, onsetDan04-1}
in one-dimensional space with periodic boundary conditions, 
\begin{equation}
\partial_t \psi(x,t)=
-\partial_{x}^{2}[\epsilon-(1+\partial_{x}^{2})^{2}]\psi-(\partial
_{x}\psi)^2. \label{onset1}
\end{equation}
Because this class of spatiotemporal chaos appears in many types of
physical systems, studying \E{1} is important 
(see, e.g., the introduction part of \cite{onsetDan04-2}). 
When we introduce a variable $v$ defined as $v \equiv 2 \partial_x \psi$, 
we can rewrite \E{onset1}: 
\begin{equation}
\partial_t v(x,t)=
-\partial_{x}^{2}[\epsilon-(1+\partial_{x}^{2})^{2}]v-v\partial
_{x}v,  \label{onset2}
\end{equation}
which we use in the following. 
This equation has the two parameters, 
the bifurcation parameter $\epsilon$ and the system size $L$. 
We know already that the spatiotemporal chaos in \E{onset2}
occurs supercritically 
when $\epsilon$ increases in sufficiently large system 
\cite{onsetTri}. 
This can be confirmed from the fact that 
the amplitudes of turbulent fluctuations vanish at the limit $\epsilon=0$ 
\cite{onsetMat, onsetDan04-2}. 
In this paper, 
we concentrate how the chaos occurs 
when the system size increases from sufficiently small size, i.e., 
we elucidate bifurcation route to chaos in \E{onset2} 
increasing the aspect ratio. 
If characteristic wavelengths of spatiotemporal patterns are comparable 
to the system size, 
then the number of active modes accommodated by the system 
is of order $1$, i.e., 
the systems behave in much the same way 
as systems of a few degrees of freedom. 
Thus, we can conjecture that 
the bifurcation route to chaos in \E{onset2} 
when the aspect ratio increases
is 
period-doubling cascade (Feigenbaum scenario) or  
intermittency transition (Pomeau-Mannville scenario) or 
breakdown of ${\rm T}^2$-torus (Ruelle-Takens-Newhouse scenario) 
\cite{onsetTuf, onsetOtt, onsetMan}. 

First, we concern, instead of \E{onset2},  
a well-known model exhibiting spatiotemporal chaos,  
Kuramoto-Sivashinsky (KS) equation 
\begin{equation}
\partial_t v(x,t)=
-\partial_{x}^{2}(1+\partial_{x}^{2})v-v\partial
_{x}v. \label{onset3}
\end{equation}
Its trivial solution $v=0$ is unstable with respect to 
perturbations having wavenumbers $q \in (0,1)$ 
and stable with respect to perturbations having the other wavenumbers.
Since the wavenumbers are, in finite-size space $L$, 
restricted to being $2\pi \nu/L$ with any integer $\nu$, 
all wavenumbers lies out of the unstable band $(0,1)$ 
if $L$ is smaller than $L_c^{KS} \equiv 2 \pi$, 
and then the uniform steady state $v=0$ is stable. 
When we increase $L$ slightly from $L_c^{KS}$, 
a few modes enter in the band $(0,1)$, i.e.,
they destabilize the uniform steady state, and 
the system exhibits a dynamical state that 
must be described as a low-dimensional attractor. 
To visualize the attractor, 
we chose a subspace ${\rm Re} A_1 - {\rm Re} A_2$, 
onto which we project the attractor,  
where $A_\nu(t)$ are the Fourier coefficients defined as 
\begin{equation}
v(x,t)=\sum_{\nu \in {\rm Z}} A_\nu (t) e^{i 2\pi \nu x /L}. \label{onset4}
\end{equation}
This choice of subspace is relatively good to see topology of the attractor.
When we increase $L$ from $L_c^{KS}$, 
the trajectory ${\bm (}{\rm Re} A_1(t), {\rm Re} A_2(t){\bm )}$    
is periodic at first as shown in \F{ribbon}. 
Then, the attractor loses its symmetry as shown in \F{asym_ribbon}. 
After that, the period of the trajectory is doubled 
as shown in \F{doubled_ribbon}. 
Finally, the attractor becomes chaotic as shown in \F{chaos_ribbon_reflex}. 
The return map on a Ponicar\'{e} section of this chaotic attractor
is unimodal, as shown in \F{unimodal}, 
which means that the KS chaos 
occurs through the period-doubling cascade 
(Feigenbaum scenario). 
In Ref.\cite{onsetK84}, Y.~Kuramoto shows 
a similar unimodal map for \E{onset3} 
with another boundary conditions $v=\partial_x^2 v=0$. 
\begin{figure}\begin{center}
\resizebox{0.44\textwidth}{!}{\includegraphics{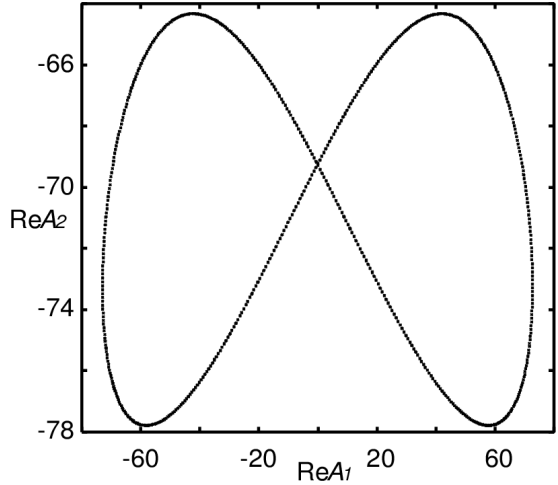}}
\caption{\label{fig:ribbon}%
Periodic attractor obtained from the KS equation with $L=2.8L_c^{KS}$. 
}
\end{center}\end{figure}
\begin{figure}\begin{center}
\resizebox{0.44\textwidth}{!}{\includegraphics{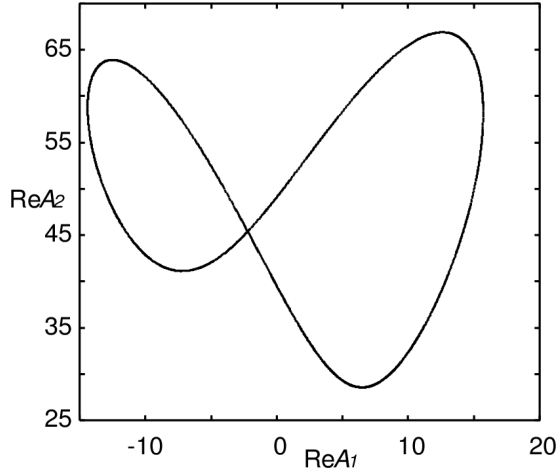}}
\caption{\label{fig:asym_ribbon}%
Symmetry-broken attractor obtained from the KS equation with $L=2.87L_c^{KS}$.
}
\end{center}\end{figure}
\begin{figure}\begin{center}
\resizebox{0.44\textwidth}{!}{\includegraphics{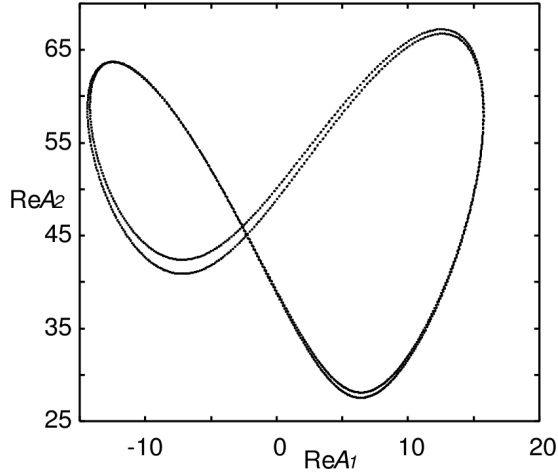}}
\caption{\label{fig:doubled_ribbon}%
Period-Doubled attractor obtained from the KS equation with $L=2.871L_c^{KS}$.
}
\end{center}\end{figure}
\begin{figure}\begin{center}
\resizebox{0.44\textwidth}{!}{\includegraphics{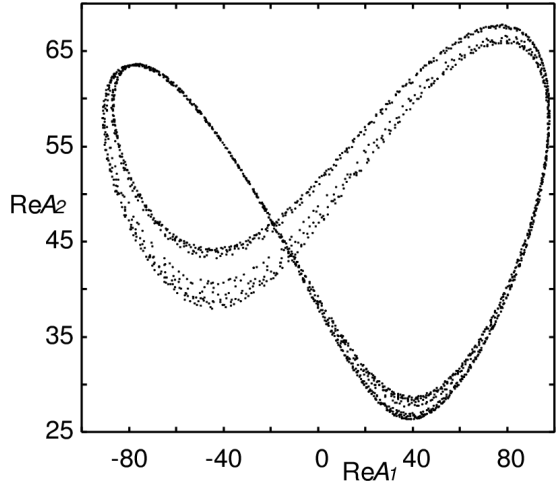}}
\caption{\label{fig:chaos_ribbon_reflex}%
Chaotic attractor obtained from the KS equation with $L=2.872L_c^{KS}$.
}
\end{center}\end{figure}
\begin{figure}\begin{center}
\resizebox{0.44\textwidth}{!}{\includegraphics{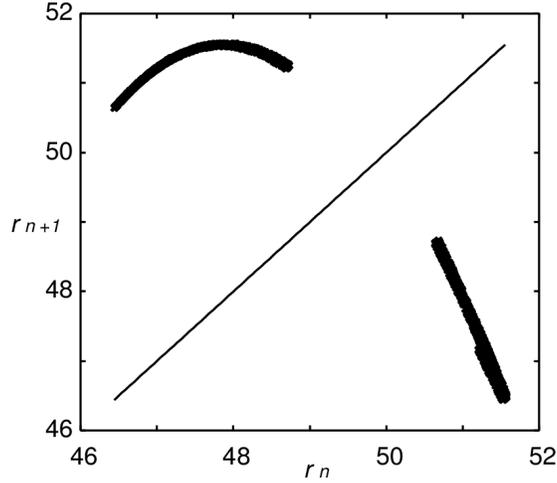}}
\caption{\label{fig:unimodal}%
Return map for the chaotic trajectory shown in \F{chaos_ribbon_reflex}. 
That trajectory crosses the Ponicar\'{e} section ${\rm Re} A_1 = 0$ 
near ${\rm Re} A_2 = 40$ and near ${\rm Re} A_2 = 50$ one after another.  
$r_n$ are defined as ${\rm Re} A_2$ at 
the intersections near ${\rm Re} A_2 = 50$. 
The KS chaos occurs through the period-doubling cascade 
(Feigenbaum scenario) 
because this map is unimodal. 
}
\end{center}\end{figure}

Now, we elucidate the onset of chaos in \E{onset2}. 
This equation also has the trivial solution $v=0$. 
Its unstable band is $(\sqrt{1-\sqrt{\epsilon}},\sqrt{1+\sqrt{\epsilon}})$. 
Thus, the critical system size is 
$L_c^N \equiv 2\pi/\sqrt{1+\sqrt{\epsilon}}$
(, which is equal to $5.4766\cdots$ when $\epsilon=0.1$). 
In the following, we use $\epsilon=0.1$.
This value is sufficiently small in order to observe 
characteristic spatiotemporal chaos exhibited 
by \E{onset2} in large systems \cite{onsetDan04-2}.
Using \E{onset4}, 
we choose a subspace ${\rm Re} A_1 - {\rm Im} A_1$, 
which is relatively good choice to see topology of the attractor.
The attractor in slightly larger system-size than $L_c^N$ 
is shown in \F{orbit}, 
where the trajectory seems to be quasi-periodic. 
In fact, 
the return map on a Ponicar\'{e} section of this trajectory 
is a closed orbit as shown in \F{secT2}, 
which means that the attractor is ${\rm T} ^2$ torus. 
When we still increase $L$, 
we see the torus doubled first, as shown in \F{secT2doubled}, 
and then chaos, as shown in \F{secCh}. 
Thus, 
the Nikolaevskii chaos occurs through the breakdown of ${\rm T}^2$-torus 
(Ruelle-Takens-Newhouse scenario). 
\begin{figure}\begin{center}
\resizebox{0.44\textwidth}{!}{\includegraphics{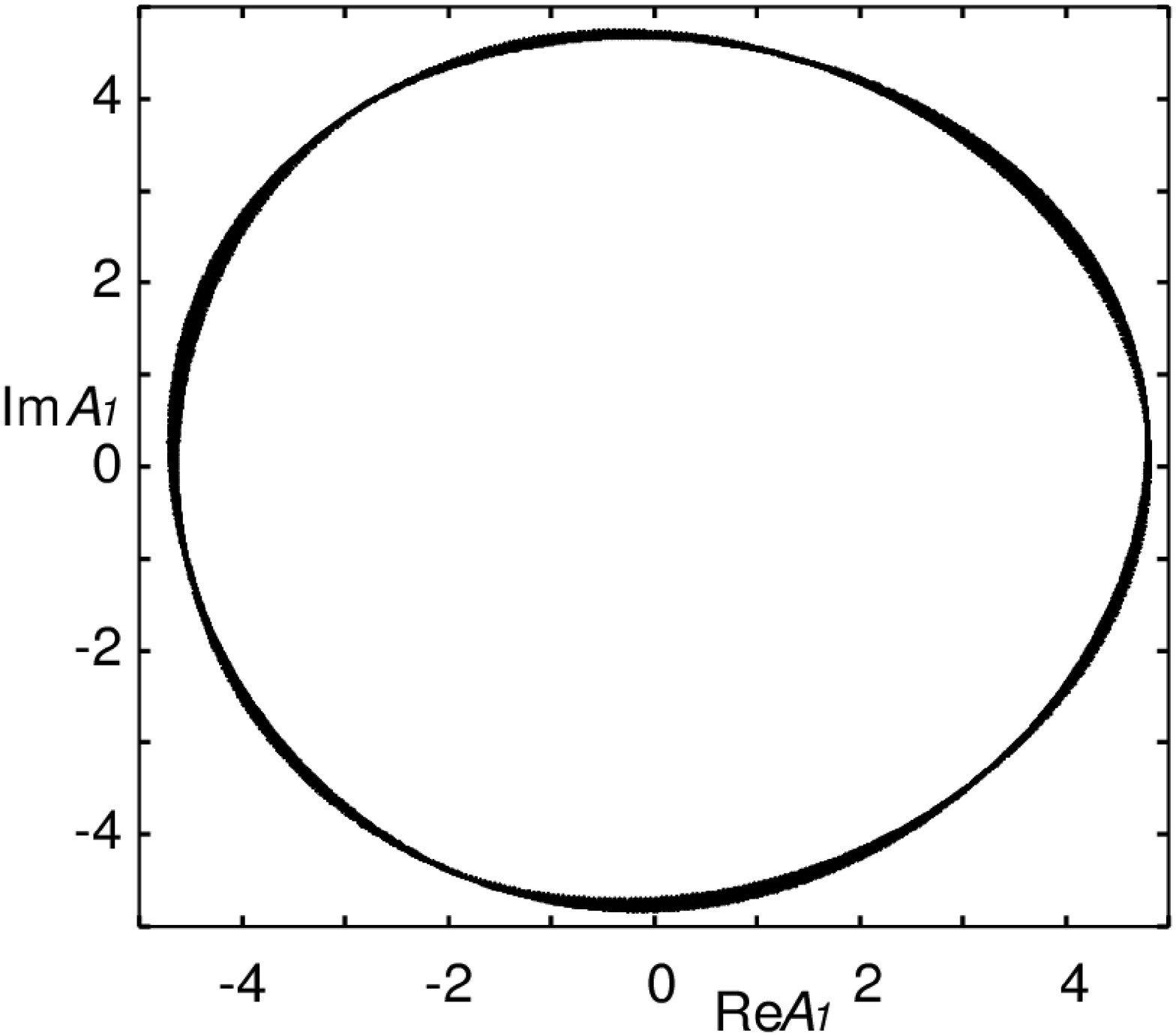}}
\caption{\label{fig:orbit}%
Quasi-periodic attractor
obtained from the Nikolaevskii equation with $L=3.97L_c^N$ and $\epsilon=0.1$. 
In this figure, the width of the orbit undulates slightly. 
However, we confirmed that this undulation disappears 
when we plot the trajectory for a longer time interval.
It means that the trajectory has two incommensurate frequencies. 
}
\end{center}\end{figure}
\begin{figure}\begin{center}
\resizebox{0.44\textwidth}{!}{\includegraphics{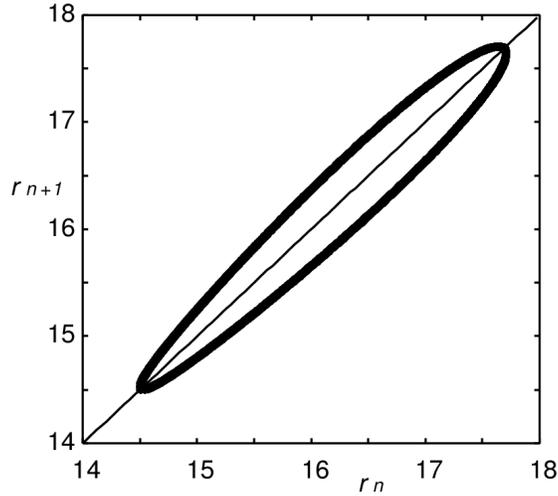}}
\caption{\label{fig:secT2}%
Return map 
obtained from the Nikolaevskii equation with $L=3.98L_c^N$ and $\epsilon=0.1$. 
$r_n$ are defined as $|A_1|$ 
on the Ponicar\'{e} section ${\rm Re} A_1 = {\rm Im} A_1 >0$.
The closed orbit is a cross section of ${\rm T} ^2$ torus. 
}
\end{center}\end{figure}
\begin{figure}\begin{center}
\resizebox{0.44\textwidth}{!}{\includegraphics{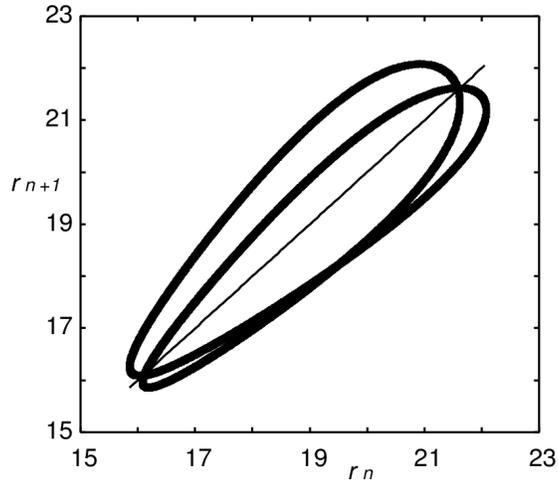}}
\caption{\label{fig:secT2doubled}%
Return map 
obtained from the Nikolaevskii equation with $L=3.9845L_c^N$ and $\epsilon=0.1$. 
The definition of $r_n$ is the same as in \F{secT2}. 
We see ${\rm T} ^2$-torus--doubling. 
}
\end{center}\end{figure}
\begin{figure}\begin{center}
\resizebox{0.44\textwidth}{!}{\includegraphics{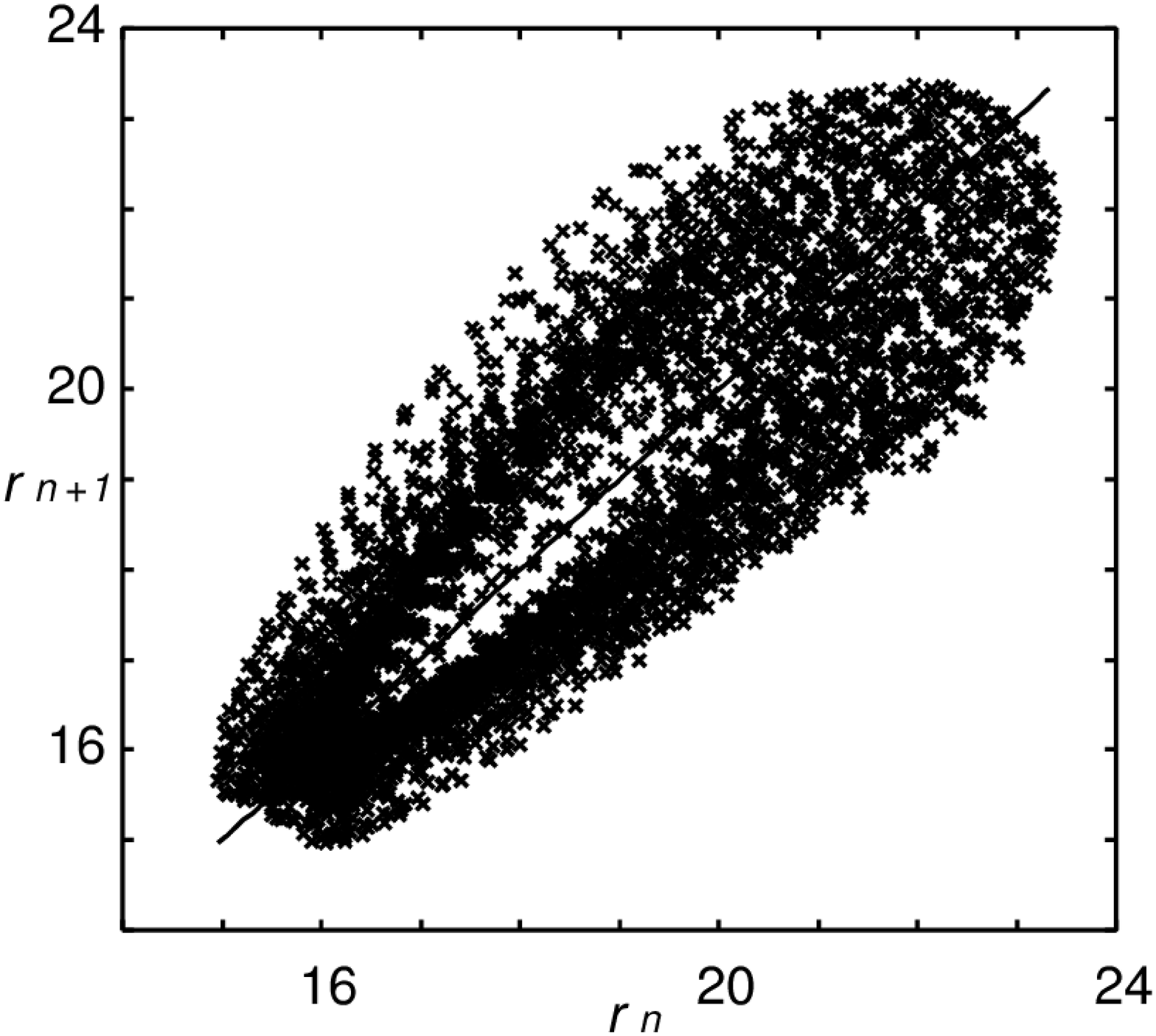}}
\caption{\label{fig:secCh}%
Return map 
obtained from the Nikolaevskii equation with $L=3.985L_c^N$ and $\epsilon=0.1$. 
The definition of $r_n$ is the same as in \F{secT2}. 
The Nikolaevskii chaos occurs through 
the ${\rm T}^2$-torus--doubling bifurcation 
(Ruelle-Takens-Newhouse scenario). 
}
\end{center}\end{figure}

Finally, we comment a future issue.
In large systems, 
the spatial power spectrum of the Nikolaevskii equation 
with $\epsilon \geq O(0.1)$ is qualitatively indistinguishable from 
that of the Kuramoto-Sivashinsky equation \cite{onsetDan04-2}. 
According to this fact, 
does the Nikolaevskii equation with $\epsilon \geq O(0.1)$ exhibit 
the Feigenbaum scenario as the Kuramoto-Sivashinsky equation ?
If so, how does 
the Ruelle-Takens-Newhouse scenario change to 
the Feigenbaum scenario ? 

D.~T.~ gratefully acknowledges 
financial support by the Japan Society for the Promotion of Science (JSPS).

\end{document}